\input{aipcheck}

\documentclass[
    ,final            
  ]
  {aipproc}

\usepackage{epsfig}
\layoutstyle{8x11single}

\newcommand{\be}{\begin{equation}}
\newcommand{\ee}{\end{equation}}
\newcommand{\bea}{\begin{eqnarray}}
\newcommand{\eea}{\end{eqnarray}}

\begin{document}

\title{Phenomenological Quantum Gravity }
\author{Dagny Kimberly}{
  address={Theory Group, Imperial College, Blackett Laboratory, Prince Consort Road,
  London SW7 2BZ}
}

\author{Jo\~{a}o Magueijo}{
  address={Theory Group, Imperial College, Blackett Laboratory, Prince Consort Road,
  London SW7 2BZ}
}

\begin{abstract}
These notes summarize a set of lectures on phenomenological quantum gravity
which one of us delivered and the other attended with great
diligence. They cover an assortment of topics on the border between
theoretical quantum gravity  and observational anomalies. Specifically, we
review non-linear relativity in its relation to loop quantum
gravity and high energy cosmic rays. Although we follow
a pedagogic approach we include an open section on unsolved problems,
presented as exercises for the student. We also review varying constant
models:
the Brans-Dicke theory, the Bekenstein varying $\alpha$ model, and several
more radical ideas. We show how they make contact with strange
high-redshift data, and perhaps other cosmological puzzles.
We conclude with a few remaining observational
puzzles which have failed to make contact with quantum gravity,
but who knows... We would like to thank Mario Novello for organizing
an excellent school in Mangaratiba, in direct competition with a very
fine beach indeed.
\end{abstract}

\maketitle


\section{Why quantum gravity?}
The subject of quantum gravity emerged as part of the unification
program that led to electromagnetism and the electroweak model.
We'd like to unify all forces of Nature. Forces other than gravity
are certainly of a quantum nature. Thus we cannot hope to have a
fully unified theory before quantizing gravity.

To come clean about it right from the start, we should stress that
there is no compelling experimental reason for quantizing gravity.
For all we know, gravity could stand alone with respect to all
other forces, and simply be exactly classical in all regimes.
There is no evidence at all that the gravitational field ever
becomes quantum~\footnote{Exercise for the student: discuss the
gravitational field of a photon undergoing the double slit
experiment. Could you collapse the wave function by measuring the
gravitational field? Lay out all possibilities in the form of
thought-experiments. You may find it interesting to make contact
with the old problem of how classical and quantum systems
interact. Then repeat the exercise with a double slit experiment
where gravitons are used instead of photons.}. Yet this hasn't
deterred a large number of physicists from devoting lifetimes to
this pursuit.

Assaults on the problem currently follow two main trends: string/M
theory~\cite{pol,stringref} and loop quantum
gravity~\cite{rovelli,carlip}. Both have merits and deficiencies,
commented extensively elsewhere. As a poor third we mention
Regge-calculus (and lattice techniques), non-commutative geometry,
and several other methods none of which has fared better or worse
than the two main strands.

This course is {\it not} about those theories. Rather it's about
the question: {\it Where might experiment fit into these
theoretical efforts of quantizing gravity?} A middle ground has
recently emerged -- phenomenological quantum gravity. The
requirements are simple: a phenomenological formalism must provide
a believable approximation limit for more sophisticated
approaches; it must also make clear contact with experimental
anomalies that don't fit into our current understanding of the
world. The following argument illustrates what we mean by this.

When physicists find themselves at a loss they often turn to dimensional
analysis. Following this simplistic philosophy we estimate the scales where
quantum gravity effects may become relevant by
building quantities with dimensions
of energy, length and time from $\hbar$ (the quantum), $c$ (relativity)
and $G$ (gravity). These are called the Planck energy $E_P$, the Planck
length $l_P$ and the Planck time $t_P$. For instance
$E_p=\sqrt{\hbar c^5/G}\approx 1.2\times 10^{19}GeV\approx 2.2\times 10^{-5}g$.
Quantum gravitational effects are expected to kick in for energies above
$E_P$ or lengths and durations smaller than $l_p$ and $t_p$.
Beware: dimensional analysis can be too naive.

We expect quantization of gravity to take the form of a theory in
which space and time are discretized. General relativity is a
theory of curved space-time. Thus, quantum gravity should quantize
not only curvature but actual durations and lengths. Here
quantization means to replace a continuum by a discrete structure.
This may be done, as a first approximation, just as Planck did in
the first quantum theory, when he simply proposed that the energy
of a harmonic oscillator would be a multiple of a fixed energy,
$\hbar \omega$.  The same could be true for space and time: space
is granular, time has an atomic structure. This is, however, an
approximation. As we know, what actually happens in quantum
mechanics is that observables are replaced by operators with a
discrete spectrum, whose eigenvalues represent the possible
outcomes of a measurement. Similarly, in loop quantum gravity,
area and volume become operators with a discrete spectrum and the
geometry becomes a quantum state (a spin foam, more specifically).
Still, we may use the ``Planck-style'' of quantization as a basis
for a phenomenological model.

The point of this simplified quantization approach is that now we
have a springboard for contact with experiment. The argument is
based on a paradox similar in flavor to the Zeno paradox in
ancient Greek philosophy (which incidentally concerned the absurd
apparent conflict between discrete and continuum). Whatever
quantum gravity may turn out to be, it is expected to agree with
special relativity when the gravitational field is weak or absent,
and also with all experiments probing the nature of space-time on
scales much larger than $l_P$ (or energy scales smaller than
$E_P$). The granules of space-time should be invisible unless we
examine these scales with a powerful ``microscope''.

This immediately gives rise to a simple question:  {\it In whose
reference frame is $l_P$ the threshold for new phenomena?} For
suppose that there is a physical length scale which measures the
size of spatial structures in quantum space-times, such as the
discrete area and volume predicted by loop quantum gravity. Then
if this scale is $l_P$ in one inertial reference frame, special
relativity suggests it will be different in another observer's
frame -- a straightforward implication of Lorentz-Fitzgerald
contraction, easily derived from the Lorentz transformations. In
other words the border between classical and quantum gravity is
not invariant or well defined. Similar arguments can be made with
energy and time.

There are two obvious answers to the problem. On the one hand,
Lorentz transformations may be correct on all scales, such that
the Planck length is sensitive to Lorentz contraction.  In this
case, quantum gravity picks up a preferred frame in which the
Planck length is the border between classical and quantum gravity.
On the other hand, it could be that quantum gravity is the ether
wind, and that all other effects baffle the Michelson-Morley
experiment.

A distinct possibility is for quantum gravity to respect the
principle of relativity, but require a revision of the Lorentz
transformations at extreme scales. Such transformations should
leave the Planck length invariant. A toy model for this is to let
the speed of light be length dependent, and go to infinity for
$l_p$. A more sophisticated version is non-linear relativity, also
called doubly special relativity (DSR), developed in the next
section. This is the prototype of the argument leading to
phenomenological quantum gravity.

The strength of this last approach is that a relation to
observations is quickly obtained in this way. Indeed, DSR explains
ultra high energy cosmic ray anomalies. This illustrates what is
meant by phenomenological quantum gravity. The theoretical problem
is too hard. Perhaps it needs a bit of fresh air, called
experiment. A simplified formalism could then be set up with the
flavors of attempts at a full solution. That is, such a formalism
can act as a target for low-energy approximations to the the full
solution, and can also make immediate contact with experiment. A
bridge between theory and experiment has been set up.

\section{Nonlinear Relativity}

"Doubly Special Relativity" (DSR) ~\cite{amelstat,gli,leejoao,leejoao1} is a
semiclassical theory, formulated in flat space-time, yet
significant at extremely high energy scales.  It is based on a
non-linear extension of the laws of Special Relativity
~\cite{leejoao}.  Just as Einstein's relativity theory is
"special" because it holds invariant the speed of light (c) as a
fundamental relativistic scale, non-linear relativity is "doubly
special" because it fixes, in addition, a relativistic {\it
energy} scale.  We stress the word "relativistic" in order to
highlight what type of fundamental scale we are dealing with.  For
instance, $\hbar$ is not a relativistic scale since it does not
affect the transformations between inertial observers like $c$
does at very high velocities. The new fixed energy scale will play
a role, independent of $c$, in relativistic transformations at
very high energies.

Part of the justification for the introduction of an invariant
energy scale into Special Relativity can be found in the lineage
of Einstein's theory. Galileo's original expression for the energy
of a fundamental particle, $E=p^2/2m$, is linearly invariant under
the transformations he defined circa 1600 A.D..  These Galilean
Transformations describe the relativity of inertial motion. They
state that position and time coordinates measured in a "primed"
frame moving at velocity $v$ in the $x$-direction with respect to
a lab frame at rest are expressed as $x'=x-vt$ and $t'=t$,
respectively. A length $\ell$, a time interval $t$ and the
velocity of a particle moving at speed
$v_1$ in the moving frame would then be written \bea v_1'&=&v_1-v \\
\ell '&=&\ell \\ t'&=&t .\eea

Roughly 200 years later Maxwell formulated his equations
describing electricity and magnetism and introduced the notion of
a constant value for the speed of light.  In order for this
invariance to hold whilst satisfying Galilean transformations, the
notion of a ``preferred observer'' had to be introduced.  The
preference was made manifest by introducing a uniform cosmic
background called the ``ether''.  In other words, the relativity of
inertial observers was lost.

While Michaelson and Morley worked to prove the constancy of the
speed of light proposed by Maxwell, Einstein's work in the early
1900s demolished the concept of the ether by unifying the
previously disjoint notions of time and space in new laws
describing the relativity of inertial motion. His discovery of the
equivalence of mass and energy led him to a revised definition of
particle energy, the quadratic "dispersion relation"
$E^2=p^2c^2+m^2c^4$.  This was no longer, however, invariant under
Galilean transformations.  In order to reestablish observer
independent laws, new relativistic transformations, the Lorentz
transformations, had to be formulated.  With respect to a frame at
rest, a particle will have coordinates in a frame moving at
velocity $v$ in the $x$-direction given by, \bea
t'&=&\gamma (t-{vx\over c^2}) \\ x'&=&\gamma (x-vt) \\ y'&=&y \\
z'&=&z , \eea where $\gamma=1/\sqrt{1-v^2/c^2}$. With respect to
energy and momenta coordinates these transformations become, \bea
E'&=&\gamma(E-vp_x)
\\ p_x'&=&\gamma(p_x-{Ev\over c^2}) \\ p_y'&=&p_y \\ p_z'&=&p_z .
\eea Galileo's linear velocity addition laws, invariant lengths
and invariant time intervals were accordingly transformed into
nonlinear velocity addition laws, length contraction and time
dilation formulae given by, \bea v_1'&=&{v_1 +v \over 1+v_1v} \\
\ell '&=&{\ell\over\gamma}
\\ \Delta t'&=&\gamma\Delta t .\eea  With these new laws governing
inertial motion, Einstein enforced the equality of all observers.

In this vein, in order to preserve the relativity of inertial
observers, DSR theories deform the Lorentz transformations such
that the modified dispersion relation,
$E^2=\textbf{p}^2c^2+m^2c^4+f(E,\textbf{p}^2;E_0)$ ~\cite{
leejoao}, is non-linearly invariant under their action. $E_0$ is
the invariant energy scale. The $f$-function in this expression
represents a generalized way of introducing the energy-dependence
into the Lorentz transformations, as we will see below.

History aside, there are two significant motivational areas for
introducing a fixed energy scale into Einstein's relativistic
transformations, one motivated by issues in quantum gravity
theories (e.g. \cite{leejoao}) and one by anomalous observations
of the cosmos (e.g. \cite{am}). Included in the first is the idea
of a fundamental energy (or length) scale, such as would arise in
a theory of quantum gravity. This scale would take on a certain
value in one frame of reference, but when boosted to another frame
of reference it would assume another value according to the
Lorentz transformations.  In fact, there would exist such a frame
in which the given energy scale would appear to surpass the
limiting energy value predicted by quantum gravity. It is thus
paradoxical for a "fundamental" scale to be relativistic, and so,
in principle, we want to modify the transformations such that they
hold this fundamental scale fixed while preserving observer
independence~\cite{amelstat,gli,leejoao}. This modification would
manifest itself only as we approach extremely high energy scales.
Otherwise, by the correspondence principle, if we set this energy
scale to infinity (or length scale to zero), which is equivalent
to not having a limiting scale at all, we recover Special
Relativity and its Lorentz transformations.  This is reflective of
the $c\rightarrow\infty$ limit taking Lorentz transformations to
Galilean transformations, and the $\hbar\rightarrow 0$ limit
taking quantum mechanics to classical mechanics.

Another theoretical motivation within the realm of quantum
gravity, is the fact that theories of Non-Commutative Geometries,
Stringy Space-Time Foams, and Loop Quantum Gravity all predict
modified dispersion relations ~\cite{leejoao}.  So, the hope would
be to come to a deformed DSR dispersion relation that is in accord
with one of these theoretical predictions.  But, naturally,
observational support is needed.  As we will show below, deformed
dispersion relations, manifest at high energy scales, can be used
to clarify anomalous observations of high energy particle
interactions.  In particular, an initial goal for DSR theories was
to explain the Ultra High Energy Cosmic Rays (UHECRs) that are
observed to collide with the Earth's atmosphere, but which are
theoretically predicted not to exist due to threshold interactions
of cosmic rays with the Cosmic Microwave Background (CMB) as they
travel through space ~\cite{review,crexp,cosmicray,leejoao1} (see
also~\cite{amel,amel1,liouv}).

Throughout these notes we will choose the Planck Energy,
$E_p\approx \sqrt{\hbar c^5/G}=10^{19}GeV$, as our fundamental energy
scale, where $\hbar$ is Planck's constant, $c$ is the speed of
light, and $G$ is Newton's gravitational constant.  This energy is
the fundamental scale for many quantum gravity theories and is
chosen as an explicit example here for simplicity.  We do not
ignore the fact that it may {\it not} be precisely this value that
should come into play in the DSR equations. The exact value must
be predicted uniquely and consistently by both quantum gravity and
observation.

\subsection{The Lorentz group}

Before launching into the construction of DSR theories we will
review some basics of the structure of the Lorentz Group, which is
an example of the general class of Lie groups, pivotal to
understanding the relativistic transformations to hand. The
"generators" of the Lorentz group actions are the infinitesimal
transformations \be L_{ab}=p_a{\partial\over\partial p^b}-
p_b{\partial\over\partial p^a},\ee where $p$ denotes the
energy-momentum 4-vector such that the lettered indexes run over
the values $(0,1,2,3)$ and Latin indexes, $i,j,etc.$ run over the
spatial indexes $(1,2,3)$. The metric signature we will use
throughout the notes is $(-,+,+,+)$. The $L_{ij}$ are rotations
about the $j$-axis, and the $L_{0j}$ are boosts in the
$j$-direction.

The commutation relations of the Lorentz generators form the {\it
algebra} of the Lorentz group.  The generators, in general, do not
commute, which allows for the algebra to be {\it closed}. This
means that the action of any two generators applied in sequence
and then replied in reverse sequence will result in an action that
is an element of the original group. Defining the boosts and
rotations to be $B_j\equiv L_{0j}$ and $R_j\equiv L_{ij}$,
respectively, the Lorentz group algebra can be written,
\bea \left[B_i,B_j\right]&=&-R_k \\ \left[R_i,R_j\right]&=&iR_k \\
\left[B_i,R_j\right]&=&iB_k \\ \left[B_i,R_i\right]&=&0 .\eea

Exponentiating these generators gives us the full finite Lorentz
transformations, $p_a'=e^{\theta L_{ab}}p_b$. Consider, for
instance, the component $L_{01}=p_0{\partial\over\partial p^1}-
p_1{\partial\over\partial p^0}$.  Exponentiating it and letting it
act on the energy-momenta coordinates will give the finite Lorentz
transformation shown above.  We do this by tailor expanding the
exponential function.  For the case of energy, $p_0$, the
expansion reads $p_0'=(1+\theta
L_{01}+{{\theta}^2\over{2!}}L_{01}^2
+{{\theta}^3\over{3!}}L_{01}^3+...)p_0$. It can be shown that the
action of $L_{01}$ on energy and momentum is given by
$L_{01}p_1=p_0$ and $L_{01}p_0=p_1$, respectively.  Thus, \bea
p_0'&=&p_0+\theta
p_1+{{\theta}^2\over{2!}}p_0+{{\theta}^3\over{3!}}p_1+...
\\ &=&p_0cosh\theta+p_1sinh\theta. \eea  Similarly,
$p_1'=p_1cosh\theta+p_0sinh\theta$.  These expressions tell us
that Lorentz transformations are rotations in hyperbolic space. We
can express the boost factor as
$\gamma={1\over\sqrt{1-v^2/c^2}}=cosh\theta$, where the velocity
is given by $v=tanh\theta$.

A final group property that we will need in the construction of
our DSR theory is a {\it realization}.  A realization of the
Lorentz group is a set of generators that has the same algebra as
the original group. It is comprised of the generators
$K_i=U^{-1}B_iU$ and $J_i=U^{-1}R_iU$, for any function $U$ such
that the Lorentz algebra relations hold. For instance, we show
\bea
[K_i,K_j]&=&U^{-1}B_iUU^{-1}B_jU-[i\leftrightarrow j] \\
&=&U^{-1}[B_i,B_j]U \\ &=&-iU^{-1}R_kU \\ &=&-iJ_k .\eea  Let us
now look explicitly at what happens to the Lorentz algebra in DSR
theory.

\subsection{Implications of a Non-Quadratic Invariant}

The DSR theory we will be discussing keeps the value of the
fundamental Planck energy scale, $E_p$, constant.  In so doing, it
preserves the Lorentz algebra, such that the DSR generators
constitute a realization of the Lorentz Group ~\cite{leejoao}.
Recall that the deformed dispersion relation introduced above,
$E^2=\textbf{p}^2c^2+m^2c^4+f(E,\textbf{p}^2;E_p)$, incorporates
an extra term representing a new dependence on a fundamental
energy, $E_p$. This function, $f$, is a map from energy-momentum
space into itself. We can thus write this as a function acting on
energy-momentum coordinates. In particular, we can write the
modified boosts of DSR as \be K_i=U^{-1}(E_p)B_iU(E_p),\ee where
$B_i$ are the special relativistic linear boost generators.  The
$E_p$-dependence now appears in the map, $U$, and must be
non-linear in order to keep $E_p$ invariant ~\cite{leejoao1} (see
also \cite{fock,man,step}). The rotation transformations in DSR
remain unchanged under this action, such that the deformed boost
generators are all that is necessary to satisfy the aims that DSR
looks to achieve ~\cite{leejoao}. We do not wish to modify the
rotations anyways since rotational invariance has been proved
accurate to extreme precision.  The transformed boosts would then
hold invariant the nonlinear expression,
$m^2=\eta^{ab}U_a(p)U_b(p)$.

 To clearly illustrate the effects that DSR has on Special
Relativity, we will introduce the $E_p$ energy scale into the
$U$-map via a particular example ~\cite{leejoao}. Consider the
deformed boost generator given by \be K_i=L_{0i}+{p_i\over E_p}D,
\ee where $D=p_a{\partial\over\partial p^a}$ is the {\it
dilatation} generator. It can be shown that this simple form
leaves $E_p$ invariant as well as the Lorentz algebra unchanged.
The corresponding $U$-map, the modification to the original
Lorentz boost generator, can thus be written \be U=e^{{E\over
E_p}D}, \ee where $E=p_0$ always. Expanding this in a tailor
series, we can find its action on the energy-momenta coordinates.
The exact expression is \be U(p_a)={p_a\over 1-{E\over E_p}},\ee
where we have a new factor in the denominator that, by the
correspondence principle, reduces to unity if we let
$E_p\longrightarrow\infty$. If, on the other hand, $E=E_p$, the
expression blows up, creating the invariant behaviour of $E_p$ we
desire.

Exponentiation of the $K_i$ generator gives us the finite form for
the transformed boosts.  However, we need not perform this
calculation. Instead, we can derive their form in a
much simpler manner by applying our $U$-map to the regular special
relativistic boost equation ~\cite{leejoao}.  That is, \bea
U(E')=\gamma(U(E)-vU(\textbf{p})) \\
U(p')=\gamma(U(\textbf{p})-vU(E)), \eea where we have set $c=1$
for simplicity.  By tailor expansion it can be proved that indeed
$e^{\theta U^{-1}B^iU}=U^{-1}e^{\theta B^i}U$.  To make this
explicit, $U$ linearizes the physical momentum, which is then
boosted by $e^{\theta B^i}$ and then converted back to a physical
momentum via $U^{-1}$: \be p_{physical}'
\stackrel{U^{-1}}{\longleftarrow}p_{linear}' \stackrel{e^{\theta
B^i}}{\longleftarrow}p_{linear}
\stackrel{U}{\longleftarrow}p_{physical} . \ee

Using the particular form for the $U$-map shown
above~\cite{leejoao}, we can solve for $E'$ on the left hand side.
This is equivalent to applying the $U^{-1}$ map, thus giving us
our deformed transformations.  These now hold invariant the
mass-squared expression
\be
m^2={\eta^{ab}p_ap_b\over (1-{E\over
E_p})^2}
\ee
For a boost in the $x$-direction, the transformations
assume the form~\cite{leejoao}, \bea E'&=&{\gamma(E-vp_x)\over
1+(\gamma-1){E\over E_p}-\gamma {vp_x\over E_p}}\nonumber\\
p'{_x}&=&{\gamma(p_x-vE)\over
1+(\gamma-1){E\over E_p}-\gamma {vp_x\over E_p} }\nonumber\\
p'{_y}&=&{p_y\over 1+(\gamma-1){E\over E_p}-\gamma {vp_x\over E_p}
}\nonumber\\
p'{_z}&=&{p_z\over 1+(\gamma-1){E\over E_p}-\gamma {vp_x\over E_p}
} .\eea  The numerators of these expressions are the familiar
special relativistic ones, whereas the denominators fully
represent the deformation.  Note that in letting
$E_p\rightarrow\infty$, we recover the special relativistic boosts
in accordance with the correspondence principle.

A simple analysis of these equations lends immediate insight into
how they are different from regular special relativistic boosts.
Most importantly, plugging $(E_p,0,0,0)$ in for $(E,p_x,p_y,p_z)$
gives $(E_p,-vE_p,0,0)$, demonstrating that $E_p$ is indeed an
invariant quantity. The regular boosts would have given us extra
$\gamma$-factors in front of the $E$ and $p_x$ values, which are
now cancelled by the new factor in the denominator of the deformed
boosts. Because of this limit, $E<E_p$ will always hold in the
single particle sector.  We will discuss the multi-particle sector
in the next section.  Similarly, rest masses will always be finite
due to this energy limit, which we can see by solving for $E$ in
the deformed dispersion relation with $\textbf{p}=0$. This gives
\be E={mc^2\over 1+{mc^2\over E_p}} ,\ee where we have reinserted
$c$ for clarity.  This expression, in turn, implies that the
classical momentum, $\textbf{p}=m\textbf{v}$, is now given by \be
\textbf{p}={m\textbf{v}\over 1+{m\over E_p}} .\ee The invariance
of $E_p$ is signalled by a singularity in the $U$-map at
$U(E_p)=\infty$. The equations have another invariant, for $E=0$,
however, this does not give the singularities necessary for an
invariant $E_p$ scale.

Now, consider the case of a Planck energy photon. Though perhaps
unphysical, this extreme case displays the limiting behavior of
these transformations.  We see that
$(E_p,E_p,0,0)\rightarrow(E_p,E_p,0,0)$.  Complete invariance.
That is to say that as $E\rightarrow E_p$, boosts become
increasingly unproductive, until ultimately, at the Planck scale,
they do no work at all.  As it turns out, this variable boost
property becomes ironically unproductive for accomplishing the
original goal of solving the UHECR anomaly ~\cite{leejoao1}, which
is discussed below.

We can also read off the new redshift formula for a photon by
setting $m=0$ in the deformed dispersion relation ~\cite{leejoao}.
This gives $E=p$, such that the boost formula gives, \be {E'\over
E}={\gamma (1-v)\over 1+(\gamma(1-v)-1){E\over E_p}}. \ee  In the
limit $E_p\longrightarrow\infty$, this reduces to the standard
Doppler redshift formula, $E'/E=\sqrt{(1-v)/(1+v)}$.  The
blueshift formula is just as above, but with the signs in front of
$v$ switched.  The deformed Doppler formula thus implies that
blueshifting a sub-Planckian photon up to $E_p$ is impossible,
because $\Delta E\longrightarrow 0$ as $E_p$ energies are
approached. This reduced Doppler effect reflects the decrease in
boost productivity as we approach Planck energies. By using the
principle of equivalence, we can get an expression for the
gravitational redshift.  Namely, \be \Delta
E/E=\Delta\phi(1-E/E_p). \ee

\section{The physical content of DSR}
We now describe how the considerations above may be used to make
contact with some anomalies and upcoming experiments. This is
expected to provide a window into quantum gravity, dependent upon
the formalism encoded in the choice of the $U$-map. There are two
main physical implications to consider.

\subsection{Ultra high energy cosmic rays}

UHECRs are believed to be highly relativistic protons that collide
with the Earth's atmosphere ~\cite{gzk}.  Due to their apparent
isotropic distribution ~\cite{alf}, they are assumed not to be of
galactic origin and thus to have traversed large distances through
the universe, interacting with the Cosmic Microwave background
(CMB) en route. There exists a theoretical bound on the energies
that the protons can have due to their threshold interactions with
the CMB photons ~\cite{gzk} . This bound, or the GZK cutoff,
occurs at a threshold energy of $6x10^{10} GeV$, due to the
$p^+\gamma_{CMB}\longrightarrow p^+\pi^0$ interaction. Note that
the cross-section for the $p^+\gamma_{CMB}\longrightarrow pee^+$
interaction does not lend a significant contribution.  The anomaly
rests in the observations of cosmic rays far above this threshold
energy ~\cite{taketal,tak} (for counter-data see ~\cite{hires}).
DSR, in altering the laws of relativistic interactions, looks to
raise the GZK cutoff in order to account for the observed ultra
high energy samples.

To see how DSR can alter the threshold energy, consider the
proton-photon interaction first in the center of mass frame such
that the proton and pion produced are at rest ~\cite{leejoao1}. In
this case, the interaction is not relativistic, so we can use
regular energy and momentum conservation laws.  These give
$E_p+E_\gamma=m_p+m_\pi$ and $E_\gamma-p_p=0$, respectively.
Boosting to the cosmological frame, the photon is redshifted to
the the temperature of the CMB radiation,  $E_\gamma'\equiv
E_{CMB}\sim2.7 K$.  The boosted proton energy is the threshold
energy, expressed $E_p'\equiv E_{th}=\gamma(E_p+vp_p)$. here we
have used the approximation $v\approx 1$ because $c=1$ in our
equations and we are dealing with a highly relativistic situation
such that $v\rightarrow c$. With these, the gamma factor can be
broken down and estimated to be $\gamma^2=1/[(1-v)(1+v)]\approx
1/[2(1-v)]$.

Using the energy and momentum conservation equations we can
rewrite the threshold energy as \be
E_{th}={(m_p+m_\pi)^2-m_p^2\over 4E_{CMB}}.\ee In general
$U_a(p)\approx p_a$ for the non-relativistic particles. But, for
the highly relativistic proton in the cosmological frame, we must
use our new DSR transformation laws, \be
E_{th}^{DSR}=U^{-1}(E_{th}^{SR}),\ee where $E_{th}^{SR}$ is the
linear threshold energy and $E_{th}^{DSR}$ is the physical
threshold energy ~\cite{leejoao1}.  Clearly, the nonlinear DSR
effects are relevant only for the energy coordinate of the
relativistic particle in the interaction, not the momenta
coordinates.

Unfortunately, the simple $U$-map used in this calculation ends up
{\it lowering} the threshold energy ~\cite{leejoao1}, which is
precisely the opposite of what we need in order to solve the
anomaly. There is, however, an alternate form of
$U$-transformation which does provide a solution ~\cite{leejoao}.
Let us first write the deformed dispersion relation in a form that
splits up the action of the $U$-map on energy and momentum
coordinates, such that
$E^2f_1^2(E,p;E_p)+\textbf{p}^2f_2^2(E,p;E_p)=m^2$ ~\cite
{leejoao1}. Choosing a $U$-map, thus corresponds to choosing the
functions, $f_1$ and $f_2$, because $U_a(p)=(Ef_1,\textbf{p}f_2)$.
In this form, we may consider particular functions that do not
necessarily act on energy and momenta in the same manner.  The
anomaly-resolving map is then
given by \be f_1={1\over(1+{E\over E_{th}})(1-{E\over E_p})} \ee
where $E_{th}$ is the energy at which the particle
threshold interaction occurs, and $f_2$ can be any function.
If you recall that the "strength",
or efficiency, of the boosts are energy-dependent in DSR, what the
introduction of this second energy scale, $E_{th}$, does is to
increase the boost efficiency around this energy in order to raise
the threshold  and thus resolve the anomaly. Above these energies
the boosts will again become increasingly unproductive as $E_p$ is
approached.

Knowing that the fundamental Planck energy scale, $E_p\sim 10^{19}
GeV$, and the cosmic ray threshold energy scale, $E_{th}\sim
10^{10} GeV$, are many orders of magnitude apart, DSR is
essentially trying to kill two birds with one stone.  It is trying
to solve the UHECR anomaly whilst holding fixed the Planck energy
scale so as to resolve the paradox of a fundamental scale being
relativistic.  Again, we mention that the particular form for the
$U$-transformation needed will be up to experiment, especially
since the UHECR data is not plentiful and conclusive
~\cite{demarc,yosh}. In particular, two caveats to consider are,
for one, the fact that the primary shower used to detect the
UHECRs is never directly seen.  Rather, it is from the analysis of
the secondary shower particles that we draw conclusions.  Thus, it
could be that the cosmic rays are not in fact protons, but some
other form of highly energetic dark matter, for instance.  Second,
further data may prove that the arrival of UHECRs is not actually
isotropic ~\cite{alf}, implying that they may originate from
within the galaxy.  Needless to say, this would crucially alter
our threshold energy calculation above, since the CMB would not
play the same role.

\subsection{Invariance of the Speed of Light and its energy dependence}
The general invariant in DSR is
\be
\eta^{ab}U_a(p)U_b(p)=m^2
\ee
where $U_a(p)=p_a^{linear}$. Given a
map of the form $U_a(p)=p_a/(1-E/E_p)$, the invariant reads
${E^2-p^2\over (1-E/E_p)^2}=m^2$. We have also just seen how we
can rewrite the dispersion relation in the even more general form,
$E^2f_1^2+\textbf{\textbf{p}}^2f_2^2=m^2$, where the new $E_p$
dependence now appears in the functions multiplying the energy and
momenta coordinates, while the rest mass, $m$, stands alone as the
new invariant.  Setting $m=0$ in the dispersion relation, we find
that the velocity for a massless particle is given by
~\cite{leejoao1} \be c(E)={E\over p}={f_2\over f_1}. \ee Note that
this is not the regular Hamiltonian expression ${dE\over dp}$.
We thus have an energy-dependent speed of light if $U$ does not
act in the same way on energy and momentum coordinates, that is if
$f_1\neq f_2$. This variability introduces a new notion of
invariance. Our invariant speed of light is now $c=c(E)$, whose
value in a boosted frame, $c(E')$, will be predicted uniquely by
the invariant form of the new deformed boosts.  What this says is
that the laws governing our universe are not fixed in the
classical sense, but rather, are allowed to change.  It is the
manner in {\it which} they change that is now invariant and
dictated by physical law.

\section{Problems for the student:
Issues Arising from Nonlinear Relativity}

We now collect a few unresolved problems within the formalism,
hoping to motivate further work.


\subsection{Fixing a U-Map}

We have said nothing so far about the reason why we have chosen
the $U$-map that we did.  The uniqueness of this map, or lack
thereof, is a key outstanding issue in DSR theories. Requesting
that $E_p$ be invariant does not point to a unique theory. That
is, $f_1$ and $f_2$ in the general deformed dispersion relation
are not uniquely specified.  This is a undesirable freedom that
cannot be resolved until further observational results are
achieved. More observations of particle threshold interactions is
just one experimental source of information for DSR.  There is
also the GLAST satellite looks to measure any energy dependence of
the speed of light ~\cite{glast}, which would lend insight into
whether the functions $f_1$ and $f_2$ are the same or not. Hence,
we leave the issue of the uniqueness of the $U$-transformation to
future experiment.

There are a couple of reasons for using the shown form for the map
~\cite{leejoao1}.  For one, it can be easily embedded in a
conformal group. Second, it is in keeping with the original
nonlinear Special Relativistic Fock-Lorentz transformations
proposed in 1964 ~\cite{fock,man}.  The difference there is that
the invariant scale in the theory was a large cosmological
distance scale as opposed to our large fundamental energy scale,
$E_p$ (equivalent to a tiny distance scale, $\ell_p$).  The map we
have been working with is also simple enough in form to enable us
to work through and concisely display results, such as time
dilation and length contraction formulae, which we will show in
the next section.

\subsection{Position Space and Field Theory}

DSR theories have always been formulated in momentum space
(however see~\cite{fock,man,step}). But, as we know, we live in
and perceive, not momentum and energy coordinates, but position
and time coordinates.  How to retrieve a position space
formulation of DSR from the existing momentum space ones, however,
is highly non-trivial.  There are two paths one can follow in
accomplishing this task ~\cite{kmm}.  The first is to begin anew
in position space, following a nonlinear special relativistic
set-up just like that which we did in momentum space.  The second
is to use the pre-existing theory in momentum space to somehow fix
the position space formulation of the transformations.

The deciding factor between these two approaches is, for the most
part, the importance of field theory.  In order to have a
kinematic theory where excitations of a field have an invariant
that correspond to that of DSR, then one must use the latter of
the two approaches above -- formulating a position space DSR
theory using the pre-existing theory in momentum space
~\cite{kmm}. To explain why, consider first the former approach of
formulating a theory directly in position space.  A map analogous
to the $U$-map used in momentum space would have to be postulated,
this time holding the Planck length, inverse to the Planck energy,
invariant.  But, this implies that we have now a two-fold freedom
from the ambiguity corresponding to the $U$-map in momentum space,
plus the ambiguity for the new map in position space, not to
mention the fact that these two maps would be completely
uncorrelated.  The fact that these maps are non-linear and would
be unrelated implies that the contraction between momentum and
position coordinates, $U^{a}(x)U_{a}(p)$, would be non-linear.

The formulation of a field theory is dependent upon the linearity
of this contraction in order for the wave solutions of a free
field theory, $\phi\sim Ae^{-ix^ap_a}$, to be planar. Maintaining
this linearity within DSR, we can use the momentum space map to
uniquely fix the position space one ~\cite{kmm}.  We can make the
standard field theoretic operator association,
$\hat{p_a}=i\partial_a$, such that
$\hat{p_a}e^{-ix^ap_a}=p_ae^{-ix^ap_a}$.  Applying this to our
deformed dispersion relation we find that the deformed Klein
Gordon Equation whose solutions are plane waves would be of the
form ~\cite{kmm} \be \left[\eta^{ab}{\partial_a\over
1-i{\partial_0\over E_p }}{\partial_b\over 1-i{\partial_0\over
E_p}}+m^2\right]\phi=0.\ee This can be proved by using Taylor
expansions of the fractional operators.

Note that this theory preserves causality with respect to the
non-linear Lorentz group because it dictates that light cones will
themselves be deformed ~\cite{kmm}.  This is due to the mixing of
space-time and energy-momentum coordinates, which we will discuss
shortly. Also, the theory does not the raise issue of
renormalizability because the theory has a natural cutoff and is
therefore strictly finite. That is, $E<E_p$ by construction, such
that there exists a frame-independent energy acting as a natural
cutoff to divergences. There exists an alternative approach,
independent of the entire formulation we have presented here,
which was proposed by Kowalski-Gilkman and uses non-commuting
space-time operators ~\cite{amelstat,gli,gama}. Our approach, on
the other hand will lead us to an energy-dependent space-time
structure ~\cite{kmm}.

Linearity of the contraction between position and momentum
coordinates ensures that the duals, $x_a$ and $p_a$, mimic one
another ~\cite{kmm}.  Relativity tells us that if $x^ap_a$ is
linear, then their transformed contraction, $U^a(x)U_a(p)$, should
be linear as well.  The map acting on position coordinates
therefore must be inverse to that acting on momentum coordinates
in order for the nonlinear factors to cancel one another.  In
general, if $U_a(p)=(Ef_1,\textbf{p}f_2)$, then
$U_a(X)=(t/f_1,\textbf{x}/f_2)$. For $U\sim e^{(E/E_p)D}$, we have
\be U_a(x)=x_a\left(1-{E\over E_p}\right).\ee  This equation
introduces a mixing of space-time coordinates with energy-momentum
coordinates due to the explicit presence of the energy, $E$.  The
corresponding invariant is energy dependent,\be
s^2=(-t^2+x^2)\left(1-{E\over E_p}\right)^2 ,\ee which shows that
we are dealing with an energy-dependent Minkowski metric.

Applying the above action onto the special relativistic boost
equation in position space gives us the deformed boosts
~\cite{kmm}, \bea t'&=&\gamma(t-vx)\left(1+(\gamma-1){E\over
E_p}-\gamma{vp_x\over E_p}\right) \\
x'&=&\gamma(x-vt)\left(1+(\gamma-1){E\over
E_p}-\gamma {vp_x\over E_p}\right) \\
y'&=&y\left(1+(\gamma-1){E\over E_p}-\gamma {vp_x\over E_p}\right)
\\ z'&=&z\left(1+(\gamma-1){E\over E_p}-\gamma {vp_x\over
E_p}\right).\eea The term that appeared in the denominator of the
deformed boosts in momentum space now appears in the numerator,
which guarantees that the linearity of the contraction of duals
holds in every inertial frame.  Again, the mixing of space-time
and energy-momentum coordinates is apparent here.

We mentioned before that light cones are deformed, allowing
causality to be preserved in DSR.  This is because of the
space-time-energy-momentum mixing occurring.  The speed of light
will be energy-dependent, but will always be the maximum velocity
and thus respect causality.  Setting $ds^2=0$ in the space-time
dispersion relation (just as we set $m=0$ in the energy-momentum
dispersion relation above) we find that the velocity for a
massless particle is given by ~\cite{kmm} \be c(E)= {dx\over
dt}={f_2\over f_1}. \ee

In order to measure or see space-time, a probe is needed, and
every probe has an energy. It is the energy of this probe that
appears in the space-time metric ~\cite{kmm}. What this is saying
is that particles with different energies will transform
differently and thus feel different space-time metrics.  We have a
"running" of geometry with energy.  The geometrical interpretation
into space-time energy-dependence is an 8-dimensional phase space
~\cite{kmm}, $(E,\textbf{p},t,\textbf{x})$, as opposed to separate
4-dimensional position and momentum spaces. What we are dealing
with geometrically is a {\it twisted bundle}. Imagine the base
space of the bundle as position space with an energy-dependent
lift into momentum space, which has a natural cutoff at $E_p$.
Near this cutoff, there is a twist in the bundle, such that the
projection down into position space is not one to one, but rather
introduces a mixing of coordinates, corresponding to deformed
topologies, or equivalently, a "running" geometry
~\cite{leejoao3}.

Before discussing the next issue in DSR, let us turn briefly to a
few familiar special relativistic formulae and what they look like
in DSR theory ~\cite{kmm}.  The addition of velocities is the same
as in Special Relativity, which can be calculated by boosting
$x=v_0t$ to get \be v'={v-v_0\over 1-vv_0} .\ee   The classic
formulae for time dilation and length contraction, however,
introduce new factors. The time dilation formula can be read off
from the deformed space-time boost equations, \be \Delta
t'=\gamma\Delta t(1+(\gamma-1){E\over E_p}), \ee where we have set
$p_x=0$ and $x=0$, consistent with a clock at the origin in the
"lab" frame. For the length contraction formula, we boost a rod of
length $L$ from an unprimed frame at rest to a primed frame moving
at velocity $v$ to get $x'=L'-vt'$,  such that \be
L'={L\over\gamma}(1+(\gamma-1){E_0\over E_p}) .\ee  $E_0={m\over
1+{m\over E_p}}$ is the particle's rest energy gotten by setting
$p=0$ in the deformed dispersion relation.  Note that the boost
parameter is equal to the origin's velocity here. For $E=E_p$ the
length contraction formula gives invariant lengths, recalling the
inefficiency of boosts that we discussed above as Planck energy
scales are approached.

\subsection{The "Soccer Ball" Problem}

The so-named "Soccer Ball Problem" is that of describing composite
systems in DSR theories.  The goal now is to extend our discussion
of the single particle sector of DSR to the multi-particle sector,
a highly non-trivial task.  With a non-linear
realization of the dispersion relation and Lorentz group, our laws
for adding momenta and energy must also be non-linear~\cite{leejoao3}.
The straightforward and classical expression,
$p_{12}=p_1+p_2$, will no longer suffice for it would immediately
imply a loss of relativity since in the boosted frame the same law
would not hold, $p_{12}'\neq p_1'+p_2'$. A simple nonlinear
composition law can be written as~\cite{leejoao3} \be
p_{12}=p_1\oplus p_2=U^{-1}(U(p_1)+U(p_2)) .\ee  In this case, the
composite momentum, $p_{12}$, is covariant, transforming as
$e^{\theta K_i}$.  This implies that the rest energy of the system
must satisfy $E_0<E_p$. But, $E_p$ corresponds to a mass of
$10^{-5}g$, which is {\it clearly} not the mass limit in our world
of composite systems. A soccer ball cannot have this energy,
except in some unusual football leagues.

An alternative and only slightly more complicated approach is
modelled by the association $U_{(N)}=U_{(1)}(E\rightarrow NE_p)$,
such that 2 particles would scale as $2E_p$ and so on up for any
number of particles. The multi particle dispersion relation, or
the momenta and energy addition laws, would then read
~\cite{leejoao3} \be {p_a^{(N)}\over 1-{E^{(N)}\over
NE_p}}=\Sigma_i{p_a^{(i)}\over 1-{E^{(i)}\over E_p}} .\ee This
addition law implies that while the system is still sub-Planckian
in the single particle sector, $E<E_p$, it can now be
super-Planckian for composites, $E_{tot}<nE_p$, as physically
expected.  The addition law is non-associative, meaning that all
particles in a system must be accounted for at once. That is, the
laws of physics would have insight into whether a system was
elementary or composite and adjust its addition laws accordingly.
In a sense, the addition laws would scale with particle number.
Note that if the particles of the system all have the same energy,
then the addition reduces to the regular linear law.

\subsection{Gravity and Kinetic Theory}
Not much to report along here. A generalisation of the idea
of an energy dependent metric for curved space-times was
proposed in~\cite{leejoao3}. Statistical physics results
assuming thermal equilibrium were first examined in~\cite{steph,steph1};
this approach may well be wrong because of the non-linearities
required to solve the soccer-ball problem invalidate most of
the partition function arguments.

These areas are still very much an exercise for the student.

\section{Varying "Constants"}

Having already glimpsed the notion of the variability of a
fundamental "constant" of nature, specifically the
energy-dependence of the speed of light as predicted by certain
formulations of DSR, we are prepared to consider other
cosmological theories where the conventional "constants" of Nature
may actually vary in space and time. There is renewed interest in
understanding the physical implications of these theories
motivated both theoretically and observationally.

Theoretically, the low energy limit of string and M quantum
gravity theories are so-called dilaton theories.  Dilaton fields,
incorporated into these quantum gravity theories for the purpose
of consistency, are massless and gauge-neutral scalar fields that
couple to matter with a strength proportional to that of the
gravitational force. Varying "constant" theories, in
which a field is used to represent a variable parameter, can be
formulated in such a way that they adopt the form of a dilaton
theory, at least in a few cases.
For parameters that are classically constant, we expect
this field to settle to a fixed value at late times -- a behavior
perhaps due to that of space-time itself.

Dirac was the first to question the constancy of the laws of
nature.  In the 1930s, he also questioned the constancy of the
constants themselves by postulating that the gravitational
parameter, $G$, may have assumed different values throughout the
evolution of the universe.  This postulate was based on the
observed evolution of the Hubble parameter, $H=\dot{a}/a$, where
$a$ is the scale factor of the universe, and its apparent, yet
unexplained, proportionality to other fundamental parameters of
nature, namely $(H_0\hbar^2/cG)^{1/3}$. He thus proclaimed that
the tiny value of $G$ today was simply a product of the old age of
the universe since his theory posited $G\propto t^{-1}$. He chose
to vary $G$ as opposed to another one of the parameters in order
not to have to reformulate atomic and nuclear physics
~\cite{wein}. The simplicity of his time varying construction made
limited predictions that ran quickly into problems. A more
successful theory was the first "varying constant" {\it field}
theory, Brans-Dicke theory, which described both temporal and
spatial variations of $G$ due to a dynamic cosmic scalar field,
$\phi$.

Variable-$G$ theories aside, Bekenstein's model of a changing
electromagnetic constant, $e$, is the next simplest theory and is
constrained not by the observational predictions of general
relativity, but rather by observations that support a variable
fine structure constant, $\alpha=e^2/4 \pi \hbar c$ ~\cite{Bek}.
Given the definition of $\alpha$, it is apparent that the
variability of $\alpha$ may be attributed not to the behavior of
$e$, but to a changing speed of light, $c$, or a variable quantum
scale, $\hbar$. Theories based on these latter two constants
~\cite{vslreview} are, however, more complex. Quantum gravity
(noncommutative geometry) in general leads to deformed dispersion
relations, which may imply a frequency-dependent speed
of light, and in turn a varying alpha value.  Since it is merely a
matter of convenience whether a varying alpha theory is formulated
in terms of a varying $e$ or $c$, the difference being in the
choice of units, there is clearly a connection drawn between the
fundamental principles of a quantum gravity theory and a varying
alpha theory, regardless of the particular choice of varying
constant. Herein also lies the connection to the DSR theories,
which are based on deformed dispersion relations, discussed above.
The simultaneous exploration of DSR and varying "constant"
theories is thus mutually beneficial in that a discovery about the
variation of constants in one theory could lend insight into the
variability determined by the other, and visa versa. Say, for
instance, that experimental data helped us determine the form of
the U-map in DSR.  This would imply whether or not the speed of
light was a constant or an energy-dependent quantity, which would
in turn guide us towards choosing the "correct" constant to vary
in the definition of alpha.

We will begin our discussion by looking at the simplest
"varying constant" theory, Brans-Dicke theory, which predicts the
variability of the gravitational constant, $G$, within the
constraints of general relativistic data.  Then we will turn to
the simplest model in the class of "varying alpha" theories, the
Bekenstein model of an electromagnetic description of the
variation of the fine structure constant, $\alpha$. Finally, we
will look at Varying Speed of Light (VSL) theories.

\subsection{Brans-Dicke Theory}

Today, the theoretical motivation for Brans-Dicke theory~\cite{bdicke}
stems, in
part, from the appearance of scalar fields coupled to gravity in
quantum gravity theories.  The hope is that the theory
would be a low energy limit of a particular quantum gravity
formulation, thus linking Planck scale physics to a notion of
testability, based on the observational evidence for variation of
the "constant", $G$. To see if such a connection is valid one must
look at the equations of motion governing the dynamics of the
scalar field, $G$. Let us look first at the classical gravity
action to understand the origin of the Brans-Dicke action and the
equations of motion derived from it.

The classical action for gravity is the Einstein-Hilbert action,
\be S_{EH}={1\over 16\pi G}\int d^4x\sqrt{-g}R .\ee  Variation of
this  with respect to the metric gives rise to Einstein's
equations of motion, \be G_{\mu\nu}=8\pi GT_{\mu\nu} ,\ee where
$G_{\mu\nu}=R_{\mu\nu}-{1\over 2}g_{\mu\nu}R$. The integrability
condition that arises in the derivation of these equations from
the variational principle is $\nabla_{\mu}G_\nu^\mu=0$ (the
so-called Bianchi identities), and
henceforth $\nabla_\mu T^\mu_\nu=0$.  The volume element is
$\sqrt{-g}d^4x$, where $\sqrt{-g}$ is invariant and $d^4x$ is the
tensor density. To get the gravitational equations of motion, we
must vary the Einstein-Hilbert action with respect to the metric,
$\delta S_{EH}=\int d^4x[(\delta(\sqrt{-g}g^{\mu\nu})R_{\mu\nu}+
\sqrt{-g}g^{\mu\nu}\delta R_{\mu\nu}]$.  Substituting
$R=g^{\mu\nu}R_{\mu\nu}$ we find a full divergence term, which can
be thrown away according to classical General relativity theory.

To calculate the equations of motion rigorously, which, for
instance, requires the calculation of $\delta
R_{\mu\nu}=\delta\Gamma^\alpha_{\mu\nu
;\alpha}-\delta\Gamma^\alpha_{\mu\alpha ; \nu}$, one must go
through the entire process of calculating $g_{\mu\nu}$, then the
connection, $\Gamma$, then the curvature tensor, $R_{\mu\nu}$, and
so on.  It is complex and actually not necessary here. Instead, we
can go to the freely falling frame where
$\Gamma^\mu_{\alpha\beta}=0$. Calculations in this frame will be
far simpler. For instance, it can be shown that $\delta
g=gg^{\alpha\beta}\delta g_{\alpha\beta}$, using the identity
$detM=e^{Tr\ell nM}$.

The general action is \be S=S_{EH}+S_{M}, \ee where $S_{M}$ is the
matter action characterized by the energy momentum tensor
$T_{\mu\nu}=-(2/\sqrt{-g})(\delta S_m/\delta g^{\mu\nu})$. Note
that we have assumed in these equations that the cosmological
constant, $\Lambda$, is zero.  If one wishes to include it, the
correct equations come from the replacement of $R$ with
$R-2\Lambda$.

We can now extend this classical theory to the case of a variable
gravitational constant.  We promote $G$ to be a function of a
scalar field, $\phi$, which produces another long-range force like
gravity. Specifically, \be \phi={1\over G} .\ee  This form comes
from the simplest generally covariant scalar field equation,
$\nabla^\rho\nabla_\rho\phi\sim T^\mu_{M\mu}$, where
$T^\mu_{M\mu}$ is the energy-momentum tensor for matter, not
including gravity and the scalar field.  A rough estimate of the
average cosmic value of $\phi$ produces a value, $10^{27} g
cm^{-3}$, close to $1/G=10^{28} g cm^{-3}$.
The gravitational action describing Brans-Dicke theory will thus be
\be S_{BD}=\int d^4x\sqrt{-g}{R\phi\over 16\pi}, \ee and the
action for $\phi$  \be S_\phi=\int
d^4x\sqrt{-g}(-{\omega\over\phi}
\partial_\mu\phi\partial^\mu\phi) ,\ee where $\omega$ is a
dimensionless coupling parameter.  The full action is the sum of
these plus the matter action, \be S=S_{BD}+S_\phi+S_M.\ee As
$\omega\longrightarrow\infty$, Brans-Dicke cosmology approaches
Einstein's gravity. The matter action is not a function of $\phi$
in this theory. Varying the Brans-Dicke action leads to a boundary
term that is no longer a full divergence as in the classical
Einstein-Hilbert action. Therefore, new terms will appear in the
field equations. Variation with respect to the metric gives \be
G_{\mu\nu}=-{8\pi\over\phi}
T_{M\mu\nu}-{\omega\over\phi^2}(\phi_{;\mu}\phi_{;\nu}- {1\over 2
}g_{\mu\nu}\phi_{;\alpha}\phi^{;\alpha})-
{1\over\phi}(\nabla_\nu\nabla_\mu\phi-g_{\mu\nu}\nabla^\mu\nabla_\mu\phi)
.\ee  The usual conservation law (i.e. the integrability
condition) holds, $\nabla_\mu T^\mu_\nu=0,$ which implies the
conservation of the standard stress-energy tensor, and henceforth
the weak equivalence principle and geodesic motion.  Variation of
the action with respect to the scalar field gives the $\phi$
equation of motion \be \nabla^\mu\nabla_\mu\phi={8\pi T\over
(3+2\omega)}, \ee where $T=T^{\mu}_{\mu}$.

Brans-Dicke Theory can be phrased in two different frames that are
conformally related.  The formulation presented above
is in the so-called {\it Jordan frame}.  The conformal, not
coordinate, transformation that takes the theory into the {\it
Einstein frame} involves the following metric replacement, \be
g_{\mu\nu\longrightarrow} g_{\mu\nu}(16\pi\phi)=\tilde{g}_{\mu\nu}
.\ee  We then reexpress the scalar field as \be \sigma={\ell
n(16\pi\phi)\over \omega+{3\over 2}} .\ee   These transformations
dress the Brans-Dicke action such that it looks like the original
Einstein-Hilbert action, solidifying the idea that Brans-Dicke
theory is just Einstein's theory with a scalar field.  An
important fact regarding this conformal transformation is that
$\phi$ does not appear in the matter action in the Jordan frame,
telling us that the scalar field does not interact with matter in
this frame. It does, however, in the Einstein frame since the
transformation $S_\sigma\rightarrow
S_M(\tilde{g}_{\mu\nu}e^\sigma)$ introduces non-minimal metric
coupling to $\sigma$. It is important to remember when making
calculations that the Einstein frame is {\it not} the physical
frame, it is merely a conformal frame in which the equations and
calculations are simplified.

Let us enter the "stringy" realm of quantum gravity for a moment.
The basic string theory action has an additional multiplicative
factor, \be S_{string}=\int d^4x\sqrt{-g}e^{-\chi}[R/16\pi+{\cal
L}_M+...], \ee where $\chi$ is a scalar field, the dilaton field
mentioned above. The form of this action displays a clear
scalar field resemblance to the Brans-Dicke action, indeed it's
its formulation in yet another conformal frame, the string frame.
The hope is
to connect varying constant theories with quantum gravity in this
way. That is, to successfully formulate a varying constant theory
that can be proven to be a limit of a quantum gravity theory.  In
this particular case, we would hope that the dilaton field was the
quantum gravitational manifestation of the $\phi$ field causing
the variability of the gravitational constant.

Finally, let us briefly address $\omega$, the new constant
introduced via Brans-Dicke theory.  It is undesirable, in general,
to introduce new parameters into a theory because they are extra
constants that need to be set or fine-tuned.  Physicists look to
constrain theories rather than introduce extra degrees of freedom.
From the Brans-Dicke equations of motion, we know that as
$\omega\longrightarrow\infty$ the theory becomes indistinguishable
from General Relativity.  In a sense, experiments that constrain
the value of $\omega$ can be seen to be a measure of just how
accurate General Relativity describes our universe given
particular experimental probes to hand.  In general, measures of
$\omega$ are quite large, with conservative estimates being in the
100s ~\cite{omega1}, and recent solar system tests using the
Cassini probe to be greater than 40,000 ~\cite{cassini}.  Such a large
range of values is not necessarily mutually exclusive, since the
tests used probe vastly different length and time scales.

There are also theories called scalar-tensor theories that allow
$\omega$ to vary along with the changing scalar field, such that
$\omega=\omega(\phi)$~\cite{fumae}. In these, the general
relativistic limit would correspond to values of $\phi$ such that
$\dot{\omega}$ is negligible and $\omega$ is very large.
%

\subsection{The Bekenstein Model}

The fine structure constant, defined as $\alpha=e^2/4 \pi \hbar
c$, is the most observationally sensitive of Nature's "constants".
 Webb, et.al. \cite{murphy,webb}, use a new observational
many-multiplet (MM) technique to measure $\alpha$, by exploiting
the extra sensitivity gained in studying relativistic transitions
to different ground states.  The tests use absorption lines from
gas clouds that intersect our line of sight to quasar (QSO)
spectra at medium redshift. This method has provided the first
evidence that the fine structure constant may change throughout
cosmological time ~\cite{murphy,webb,webb2}. Independent samples
yielded imply that the value of $\alpha$ was lower in the past,
such that $\Delta \alpha /\alpha =-0.72\pm 0.18\times 10^{-5}$ for
$z\approx 0.5-3.5$.  VLT tests of Iron absorption lines at a
redshift of $z=1.15$ back-lit by  a QSO and using the standard MM
technique result in $10^{-6}$ ~\cite{vlt}. Another test for the
variability of alpha comes from the Oklo natural fission reactor,
a uranium mine in Africa, which determines historical values of
alpha by measuring isotopes 2000 million years in age. This has
set tight constraints on the temporal variation of alpha of the
order $10^{-7}$ ~\cite{oklo}.

Big Bang Nucleosynthesis (BBN) is one of the great experimental
successes in cosmology.  It describes a phase of the early
universe when particles could coalesce to form atoms due to a
sufficient decrease in the temperature of the universe.  The
theory successfully predicts the abundances of light elements in
our cosmos today. These abundances depend critically on the number
of baryons (neutrons and protons) at the time of nucleosynthesis,
which in turn is governed by the particle interaction mediator,
the fine structure constant.   The idea here is to accommodate the
data in support of variable fine structure constant while staying
in agreement BBN successes such that we are able to fix the value
of alpha to what it is measured to be today. Some investigations
into the variability of this "constant" have come to conclude
$\Delta \alpha <0$ values in order to best fit Cosmic Microwave
Background (CMB) data at $z\approx 10^3$ and BBN data at $z\approx
10^{10}$, respectively ~\cite{avelino,bat}.

To address the question of what has made the value of alpha
increase throughout cosmological time, one must first ask, to
which of the "constants" defining alpha should we attribute the
variation? Through which parameter the variation is realized is a
matter of convenience, but it is this choice that is crucial to
formulating a simple and cohesive theory.  The most conservative
of the varying alpha theories is one in which the electromagnetic
charge, $e$, varies in space-time, while $c$ and $\hbar$ are held
fixed ~\cite{Bek}. A varying $e$ theory can be set up in much the
same way as in the Brans-Dicke case, by prescribing that $e$
become a dynamic field, the so-called minimal coupling
prescription ~\cite{bek2}. This electromagnetic varying $e$ theory
has been thoroughly explored, including a formal rearrangement of
these theories done to convert them into dilaton-type theories, in
which the dilaton couples to the electromagnetic ``$ F^2$'' term
in the Lagrangian, but not to the other gauge fields~\cite{Bek}.

In the varying $\alpha$ theories proposed initially in ~\cite{Bek}
one takes $c$ and $\hbar $ to be constants and attributes
variations in $\alpha$ to changes in $e$, the permittivity of free
space. This is done by letting $e$ take on the value of a real
scalar field which varies in space and time, $e_0\rightarrow
e(x^\mu)=e_0\epsilon (x^\mu ),$ where $\epsilon(x^\mu)$ is a
dimensionless scalar field and $e_0$ is a constant denoting the
present value of $e(x^\mu)$. One then proceeds to set up a theory
based on the principles of local gauge invariance, causality of
electromagnetism, and the scale invariance of the
$\epsilon$-field.

Since covariant derivatives in electromagnetic theory take the
form $D_\mu \phi=(\partial_\mu +ieA_\mu)\phi$, for gauge
transformations of the form $\delta\phi=-i\chi\phi$ one must
impose ~\cite{Bek} \be\epsilon A_\mu \rightarrow \epsilon A_\mu
+\chi _{,\mu }.\ee The gauge-invariant electromagnetic field
tensor is then \be F_{\mu \nu }={ (\epsilon A_\nu )_{,\mu
}-(\epsilon A_\mu )_{,\nu }\over \epsilon} , \ee which reduces to
the usual form for constant $\epsilon $. The electromagnetic
Lagrangian is still \be{\cal L} _{em}=-{F^{\mu \nu }F_{\mu \nu
}\over 4}, \ee and the dynamics of the $\epsilon $ field are
controlled by the kinetic term \be{\cal L}_\epsilon =-\frac 12
\omega {\epsilon _{,\mu }\epsilon ^{,\mu }\over {\epsilon ^2}},
\ee where the self-coupling constant $\omega$ is introduced into
the lagrangian density for dimensional reasons and is proportional
to the inverse square of the characteristic length scale of the
theory, $\omega\sim\ell^{-2}$, such that $\ell\geq
L_p\approx10^{-33}cm$ holds ~\cite{bek2}. This length scale
corresponds to an energy scale ${\hbar c \over \ell}$, with an
upper bound set by experiment.  Note that the metric signature
used is still $(-,+,+,+)$.

A simpler formulation of this theory ~\cite{bsm,joaohaav} can be
constructed by defining an auxiliary gauge potential $a_\mu \equiv
\epsilon A_\mu ,$ and field tensor $f_{\mu \nu }\equiv\epsilon
F_{\mu \nu }=\partial _\mu a_\nu -\partial _\nu a_\mu .$  We are
free to do this without physical implications because, at the
classical level, it is not $A_\mu$ that we measure, but rather the
electric and magnetic fields, $E$ and $B$.  The covariant
derivative then assumes the familiar form, $D_\mu =\partial _\mu
+ie_0a_\mu $, and the dependence of the Lagrangian on $\epsilon $
then occurs only in the kinetic term for $\epsilon $ and in the
$F^2=f^2/\epsilon ^2$ term, not in a kinetic term, $D_\mu \phi
D^\mu \phi$, of the matter lagrangian, ${\cal L}_M$. To simplify
further, we can redefine the variable, $\epsilon \rightarrow \psi
\equiv ln\epsilon .$ The total action then becomes \be
S=\int d^4x\sqrt{-g}\left({\cal L}_{M}+{\cal L}_\psi +{\cal L}%
_{em}e^{-2\psi }\right) ,\ee with \bea
{\cal L}_\psi &=&-{\frac \omega 2}\partial _\mu \psi \partial ^\mu \psi ,\\
{\cal L}_{em}&=&-\frac 14f_{\mu \nu }f^{\mu \nu }, \eea where the
matter Lagrangian ${\cal L}_{M}$ does not contain $\psi$. This is
a dilaton theory coupling to the electromagnetic ``$f^2$'' part of
the Lagrangian only ~\cite{bsm,joaohaav}.  Note that the scale
invariance of the $\epsilon$ field is such that the action is
invariant under the transformation $\epsilon \rightarrow k\epsilon
$ for any constant $k$.  The action, written in this simplified
form, clearly resembles the dilaton-string action shown above.

Minimizing the variation of the action with respect to the scalar
field, $\delta S/\delta\phi=0$, gives us the equations of motion
governing the temporal and spatial development of $\phi$
~\cite{bsm}, \be
\nabla^\mu\nabla_\mu\phi={2\over\omega}e^{-2\psi}{\cal L}_{EM}
.\ee  Ignoring, for the moment, any spatial variation in $\phi$,
we have the Friedman-Robertson-Walker equations governing the
time-evolution of $\phi$ ~\cite{bsm}, \be
\ddot{\phi}+3{\dot{a}\over
a}\dot{\phi}={-2\over\omega}e^{-2\psi}{\cal L}_{EM} ,\ee where the
right hand side of the equation is the driving force and
$(3\dot{a}/a)\dot{\phi}$ on the left hand side is the fluid
friction term. It is the electromagnetic lagrangian,
$\textbf{E}^2-\textbf{B}^2$, therefore, which drives the variation
in $e$. In a radiation-dominated universe
$\textbf{E}^2-\textbf{B}^2=0$ such that there is no variation in
$e$ and thus $\alpha$. Numerical results ~\cite{bsm} support a
variable alpha with the following behavior. The radiation era sees
no variation in $\alpha$, but the driving term is not zero during
the matter era, $\textbf{E}^2-\textbf{B}^2\neq 0$, allowing for
small variations during this period. In the ensuing epoch of
accelerated expansion the fluid friction term dominates, leading
to the stability of the field $\psi$, and hence constant $e$ and
$\alpha$.

Unfortunately, the variation in alpha predicted by these equations
with a positive energy density for the scalar field (i.e.
$\omega>0$) is precisely opposite to that needed to support
observational data.  That is, the equations predict a value of
$\alpha$ that {\it decreases} in time, whereas observations
support an {\it increasing} alpha ~\cite{bsm}. The
theory can fit the data, however, for two very different physical
scenarios. Firstly, if there exists a type of dark matter such
that the $B^2$ contribution to the electromagnetic lagrangian
dominates the $E^2$ contribution, then $\alpha$ will increase in
time within observational bounds. Some cosmological defect
theories, specifically those of superconducting strings, satisfy
this requirement, but it is more or less unappealing since most
matter types do not have this property.

It is, on the other hand, possible to explain the observational
data supporting an $\alpha$ value that was lower in the past if
$\omega<0$~\cite{bsm}. A negative scalar coupling
implies a negative energy density for $\psi$, which in turn tells
us that $\psi$ is a "ghost" field. "Ghost fields" coupled to
matter have often been deemed undesirable since they will
consistently dump positive energy into matter.  However, it is the
specific forms of the equations of motion that will dictate just
how this interchange is mediated such that it may, in fact, not be
problematic. We will ignore the "runaway" pathological behavior of
ghosts at the quantum level on the grounds that a scalar field
$\phi$ is non-renormalizable and should not be quantized, just as
in classical general relativity.

It has been shown ~\cite{bkj} that the simplest varying constant
theories, such as the Brans-Dicke and Bekenstein models described
above, predict a non-singular and cyclic universe given a scalar
coupling $\omega<0$.  In particular, the Bekenstein model
equations with negative $\omega$ are the same as the Brans-Dicke
equations in the Einstein frame. In both cases, a regularly
oscillating universe results if the scalar field is uncoupled to
matter.  On the other hand, a cyclic universe with "bounces" of
ever-increasing amplitudes is produced if the field {\it is}
coupled to matter near the bounce, such that positive energy is
transferred into the matter field from the negative energy field.
The universe, with this additional positive energy density will be
hotter and will thus have to grow to a larger size in order to
reach the critical temperature for a turnaround. Note that these
models occur strictly within the radiation-dominated epoch,
meaning that the scalar fields couple to radiation. Assuming that
we are presently in the first matter-dominated era, we can
calculate the number of bounces needed to get to the value of the
constants that are measured today ~\cite{bkj}.  In this manner, it
is possible to justify the extreme values of the parameters simply
given the large age of our universe.

We have so far reviewed a large number of motivations for varying
"constant" scenarios.  Among them are included quantum gravity
theories with dilaton scalar fields resembling the scalar fields
that describe varying constants, theories such as DSR which
predict deformed dispersion relations, and observations pointing
to the variability of the fine structure constant.  In all cases,
scalar fields play the pivotal role of describing how these
fundamental parameters of nature vary throughout space-time. We
know that scalar fields appear in other roles in physics, such as
the cosmological constant that drives inflation, so it
is more than relevant to ask what the crucial role of scalar
fields in the universe is in general.  Is their primary role to
vary the fundamental constants of nature?  Are they there simply
to source inflation?  Or, are they merely an unobservable
byproduct of higher dimensional theories such as string theories?
Perhaps these theories are intimately connected, but to even begin
to address these questions, one must broaden the scope of study.
For instance, an extension of the Bekenstein model to the
electroweak scenario been done ~\cite{varsm} in order to see if
there exists any evidence of a scalar field affecting the standard
model of particle physics. With respect to cyclic universes, there
are other non-singular models ~\cite{rob,varun} sourced other than
by negative energy scalar fields, which could help shed light on
the necessity and/or importance of scalar fields in bouncing
models.

\subsection{A Variable Speed of Light}
Even after the proposal of special relativity in 1905 many varying
speed of light theories were considered, most notably by Einstein
himself~\cite{einsvsl}. VSL was then rediscovered and forgotten on
several occasions. For instance, in the 1930s, VSL was used as an
alternate explanation for the cosmological redshift
\cite{stew,buc,wald} (these theories conflict with fine structure
observations). None of these efforts relate to recent VSL
theories, which are firmly entrenched in the successes (and
remaining failures) of the hot big bang theory of the universe. In
this sense the first ``modern'' VSL theory was J. W. Moffat's
ground breaking paper \cite{moffat93}, where spontaneous symmetry
breaking of Lorentz symmetry leads to VSL and an elegant solution
to the horizon problem.

Since then there has been a growing literature on the subject,
with several groups working on different aspects of VSL (for a
comprehensive review see~\cite{vslreview}).
Here we examine the simplest and most conservative  implementations
currently being considered. The first resembles Brans-Dicke theory
in that two metrics are considered, one for gravity,
another for matter (this is comparable to the metrics used in the
Einstein and Jordan frame in Brans-Dicke theory).
The second implementation is a generalization of Bekenstein's
theory in which all coupling strengths become dynamical.

\subsubsection {Bimetric VSL theories}\label{bivsl}
This approach was initially proposed by J. W. Moffat and M. A. Clayton
\cite{mofclay}, and by I. Drummond \cite{drum}. It does not sacrifice
the principle of relativity and special care is taken with the damage
caused to the second principle of special relativity.
In these theories the speeds of the various
massless species may be different, but special relativity is still
(linearly) realized within each sector. Typically the speed of the graviton
is taken to be different from that of massless matter particles.
This is implemented by introducing two metrics (or tetrads in the
formalism of \cite{drum,drum1}), one for gravity and one for
matter. The model was further studied by \cite{clay1}
(scalar-tensor model), \cite{clay} (vector model), and
\cite{covvsl,bass,bass1}.

We now sketch the scalar-tensor model. It uses a scalar field
$\phi$ that is minimally coupled to a gravitational field
described by the metric $g_{\mu\nu}$. However the matter couples
to a different metric, given by
\begin{equation}
\hat{g}_{\mu\nu}=g_{\mu\nu}+B\partial_\mu\phi\partial_\nu\phi.
\end{equation}
Thus there is a space-time, or graviton metric $g_{\mu\nu}$, and a
matter metric $\hat{g}_{\mu\nu}$. The total action is
\begin{equation}
S=S_g+S_{\phi}+\hat{S}_{\rm M},
\end{equation}
where the gravitational action is as usual
\begin{equation}
S_g=\frac{1}{16\pi }\int dx^4{\sqrt {-g}} (R(g)-2\Lambda),
\end{equation}
(notice that the cosmological constant $\Lambda$ could also,
non-equivalently, appear as part of the matter action). The scalar
field action is
\begin{equation}
S_\phi=\frac{1}{16\pi }\int dx^4{\sqrt {-g}}\,
\Bigl[\frac{1}{2}g^{\mu\nu}\partial_\mu\phi\partial_\nu\phi-V(\phi)\Bigr],
\end{equation}
leading to the the stress-energy tensor,
\begin{equation}
T^{\mu\nu}_\phi=  \frac{1}{16\pi } \Bigl[
g^{\mu\alpha}g^{\nu\beta}\partial_\alpha\phi\partial_\beta\phi
-\frac{1}{2}g^{\mu\nu}g^{\alpha\beta}\partial_\alpha\phi\partial_\beta\phi
+g^{\mu\nu}V(\phi) \Bigr].
\end{equation}
The matter action is then written as usual, but using the metric
${\hat g}_{\mu\nu}$. Variation with respect to $g_{\mu\nu}$ leads
to the gravitational field equations,
\begin{equation}
G^{\mu\nu}=\Lambda g^{\mu\nu}
 +8\pi   T^{\mu\nu}_\phi
 +8\pi     \frac{\sqrt{-\hat{g}}}{\sqrt {-g}}\hat{T}^{\mu\nu}.
\end{equation}
In this theory the speed of light is not preset, but becomes a
dynamical variable predicted by a special wave equation
\begin{equation}\label{dynabivsl}
\bar{g}^{\mu\nu}\hat{\nabla}_\mu\hat{\nabla}_\nu\phi+KV^\prime
[\phi]=0\,
\end{equation}
where the biscalar metric $\bar g$ is defined in \cite{mofclay}.

This model not only predicts a varying speed of light (if the
speed of the graviton is assumed to be constant), but also allows
solutions with a de Sitter phase that provides sufficient
inflation to solve the horizon and flatness problems. This is
achieved without the addition of a potential for the scalar field.
The model has also been used as an alternative explanation for the
dark matter \cite{drum1} and dark
energy\cite{bass,bass1}.

\subsubsection{``Lorentz invariant'' VSL theories}\label{livsl}
It is also possible to preserve the essence of Lorentz invariance
in its totality and still have a space-time (as opposed to energy
dependent) varying $c$. One possibility is that Lorentz invariance
is spontaneously broken, as proposed by J. W. Moffat~\cite{moffat93,moff2}
(see also \cite{jacobson}). Here the
full theory is endowed with exact local Lorentz symmetry; however
the vacuum fails to exhibit this symmetry. For example an $O(3,1)$
scalar field $\phi^a$ (with $a=0,1,2,3$) could acquire a time-like
vacuum expectation value (VEV), providing a preferred frame and
spontaneously breaking local Lorentz invariance to $O(3)$
(rotational invariance). Such a VEV would act as the preferred
vector $u^a=\phi^a_0$; however the
full theory would still be locally Lorentz invariant. Typically in
this scenario  the speed of light undergoes a first or second
order phase transition to a value more than 30 orders of magnitude
smaller, corresponding to the presently measured speed of light.
Interestingly, before the phase transition the entropy of the
universe is reduced by many orders of magnitude, but afterwards
the radiation density and entropy of the universe vastly increase.
Thus the entropy increase follows the arrow of time determined by
the spontaneously broken direction of the timelike VEV $\phi^a_0$.
This solves the enigma of the arrow of time and the second law of
thermodynamics.

Another example is the covariant and locally Lorentz invariant
theory proposed in \cite{covvsl}. In that work definitions were
proposed for covariance and local Lorentz invariance that remain
applicable when the speed of light $c$ is allowed to vary. They
have the merit of retaining only those aspects of the usual
definitions which are invariant under unit transformations, and
which can therefore legitimately represent the outcome of an
experiment. In the
simplest case a scalar field is then defined $\psi=\log( c/ c_0)$,
and minimal coupling to matter requires that \be\label{alphan}
\alpha_i\propto g_i\propto \hbar c\propto c^{q} \ee  with $q$ a
parameter of the theory. The action may be taken to be
\begin{equation} S= \int d^4x \sqrt{-g}(
e^{a\psi}( R-2\Lambda +{\cal L}_{\psi}) +{ 16\pi }e^{b\psi}{\cal L}_m )
\end{equation}
and the simplest dynamics for $\psi$ derives from:
\begin{equation} {\cal L}_{\psi}=-\kappa(\psi)
\nabla_\mu\psi\nabla^\mu\psi
\end{equation}
where $\kappa(\psi)$ is a dimensionless coupling function. For
$a=4$, $b=0$, this theory is nothing but a unit transformation
applied to Brans-Dicke theory. More generally, it's only when
$b+q=0$ that these theories are scalar-tensor theories in
disguise. In all other cases it has been shown that a unit
transformation may always be found such that $c$ is a constant but
then the dynamics of the theory becomes much more complicated.
Thus we should label these theories varying speed of light theories.

In these theories the cosmological constant $\Lambda$ may depend
on $c$, and so act as a potential driving $\psi$. Since the vacuum
energy usually scales like $c^4$ we may take $\Lambda\propto
(c/c_0)^n=e^{n\psi}$ with $n$ an integer. In this case, if we set
$a=b=0$ the dynamical equation for $\psi$ is :
\be\label{dynacovvsl} \nabla_\mu\nabla^\mu \psi ={32\pi \over c^4\kappa}{\cal
L}_m +{1\over \kappa} n\Lambda \ee Thus it is possible
that the presence of Lambda drives changes in the speed
of light, a matter examined (in another context) in \cite{vslsn}.

Particle production and second quantization for this model has
been discussed in \cite{covvsl}. Black hole solutions were also
extensively studied \cite{vslbh}. Predictions for the classical
tests of relativity (gravitational light deflection, gravitational
redshift, radar echo delay, and the precession of the perihelion
of Mercury) were also shown to differ distinctly from their
Einstein counterparts, while still evading experimental
constraints~\cite{vslbh}. Other interesting results were the
discovery of Fock-Lorentz space-time\cite{man,step} as the
``free'' solution, and fast-tracks (tubes where the speed of light
is much higher) as solutions driven by cosmic
strings~\cite{covvsl}.

Beautiful as these two theories may be, their application to
cosmology is somewhat cumbersome.

\subsubsection{The simplistic cosmological motivation}
Like inflation~\cite{infl}, modern VSL theories were  motivated by
the ``cosmological problems'' -- the flatness, entropy,
homogeneity, isotropy  and cosmological constant problems of Big
Bang cosmology (see~\cite{am,basker}). The definition
of a cosmological arrow of time was also a strong consideration.

\begin{figure}
\centerline{\psfig{file=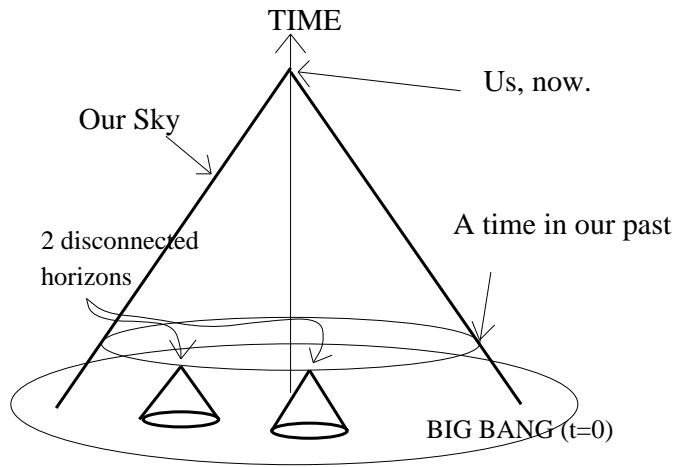,width=6 cm,angle=-90}}
\caption{A conformal diagram (in which light travels at $45^\circ$).
This diagram reveals that the sky is a cone in 4-dimensional
space-time. When we look far away we look into the past; there is
an horizon because we can only look as far away as the Universe is
old. The fact that the horizon is very small in the very early
Universe, means that we can now see regions in our sky outside
each others' horizon. This is the horizon problem of standard Big
Bang cosmology.} \label{fig1}
\end{figure}

\begin{figure}
\centerline{\psfig{file=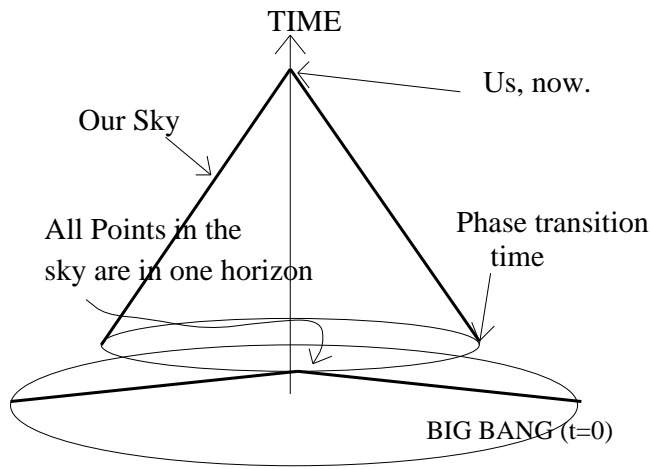,width=6 cm,angle=-90}}
\caption{Diagram showing the horizon structure in a model in which
at time $t_c$ the speed of light changed from $c^-$ to $c^+\ll
c^-$. Light travels at $45^\circ$ after $t_c$ but it travels at a
much smaller angle to the spatial axis before $t_c$. Hence it is
possible for the horizon at $t_c$ to be much larger than the
portion of the Universe at $t_c$ intersecting our past light cone.
All regions in our past have then always been in causal contact.
This is the VSL solution of the horizon problem.} \label{fig2}
\end{figure}

But at its most simplistic, VSL was inspired by the horizon problem.
As we go back into our past the present comoving horizon breaks down
into more and more comoving causally connected regions. These
disconnected early days of the Universe prevent a physical
explanation for the large scale features we observe -- the ``horizon
problem''. It does
not take much to see that a larger speed of light in the early
universe could open up the horizons~\cite{moffat93,am} (see
Figs.~\ref{fig1} and \ref{fig2}). More mathematically,
the comoving horizon is given by
$r_h=c/\dot a$, so that a solution to the horizon problem requires
that in our past $r_h$ must have decreased in order to causally
connect the large region we can see nowadays. Thus
\begin{equation}
{\ddot a\over \dot a}-{\dot c\over c}>0
\end{equation}
that is, either we have accelerated expansion (inflation), or a
decreasing speed of light, or a combination of both. This argument
is far from general: a contraction period ($\dot a<0$, as in the
bouncing universe), or a static start for the universe ($\dot
a=0$) are examples of exceptions to this rule.

However the horizon problem is just a warm up for the other
problems.
More recently structure formation has been the leading driving
force in these scenarios. This is still very much unaccomplished,
in spite of recent efforts and consequently is left as an exercise
for the student.

\subsubsection{The experimental front of VSL}

As explained in~\cite{vslreview} varying $e$ and $c$ theories in
general predict the same $\alpha(z)$ profiles.
To distinguish between varying $e$ and varying $c$ theories
one must look elsewhere. The status of the equivalence principle
in these theories turns out to be a good solution. The varying $e$
theories \cite{bsm,bek2,olive} violate the weak equivalence
principle, whereas VSL theories do not\cite{mofwep,mbswep}. The
E\"otv\"os parameter
\begin{equation}
\eta \equiv {\frac{2|a_1-a_2|}{a_1+a_2}}
\end{equation}
is of the order $10^{-13}$ in varying $e$ theories, just an order
of magnitude below existing experimental bounds.

Still, it would be good to find a widely independent confirmation
of the Webb results.
A very promising area is cold atom clocks~\cite{sortais}. The
search for the perfect unit of time has led to the the quest for
very stable oscillatory systems, leading to a gain, every ten years, of
about one order of magnitude in timing accuracy.
Cold atom clocks may be used as laboratory ``table-top''
probes for varying $\alpha$, with a current sensitivity
of about $10^{-15}$ per year.  For all varying alpha theories
it is found that at present:
\be
{\dot\alpha\over\alpha}\approx 2.98\times 10^{-16} h \quad {\rm year}^{-1}
\ee
with $H_0=100 h$~Km~sec$^{-1}$~Mpc$^{-1}$, $\Omega_\Lambda=0.71$
and $\Omega_m=0.29$. For $h=0.7$ this
gives a fractional variation in alpha of about $2\times 10^{-16}$ per year,
which should soon be within the reach of technology.
Such an observation would be
an incredible vindication of the Webb results. On the other
hand this effect would become a further annoyance for those
concerned with  the practicalities of defining
the unit of time.

Spatial variations of $\alpha $ are likely to be significant
\cite{mbswep} in any varying alpha theory. For any causal theory
relative variations in
$\alpha $ near a star are proportional to the local gravitational
potential. The exact relation between the change in $\alpha $ with
redshift and in space (near massive objects) is model dependent.
For instance, we have
\begin{equation}
{\frac{\delta \alpha }\alpha }=-{\frac{\zeta _s}\omega }{\frac{M_s}{\pi r}}%
\approx 2\times 10^{-4}{\frac{\zeta _s}{\zeta _m}}{\frac{M_s}{\pi
r}} \label{alphar}
\end{equation}
for a typical varying $e$ theory, but
\begin{equation}
{\frac{\delta \alpha }\alpha }=-{\frac{bq}\omega }{\frac{M_s}{4\pi r}}%
\approx 2\times 10^{-4}{\frac{M_s}{\pi r}}\;,  \label{alphar1}
\end{equation}
for a VSL theory. Here $M_s$ is
the mass of the compact object, $r$ is its radius, and $\zeta$
is the ratio between $E^2-B^2$ and $E^2+B^2$. When $\zeta _m$ (for
the dark matter) and $\zeta _s$ (for, say, a star) have different
signs, for a cosmologically {\it increasing}
$\alpha $, varying $e$ theories predict that $%
\alpha $ should {\it decrease} on approach to a massive object.
And indeed one must have $\zeta _m<0$ in order to fit the Webb
results. In VSL, on the contrary, $\alpha $ {\it increases}
near compact objects (with decreasing $c$ if $q<0$, and increasing $c$ if $%
q>0$). In VSL theories, near a black hole $\alpha $ could become
much larger than 1, so that electromagnetism would become
non-perturbative with dramatic consequences for particle physics near
black holes. In varying-$e$ theories precisely the opposite
happens: electromagnetism switches off.

These effects are in principle observable using similar
spectroscopic techniques to those of Webb, but applied to lines
formed on the surface of very massive objects near us (in the
sense of $z\ll 1$). For that, we
need an object with a radius sufficiently close to its
Schwarzchild radius, such as an AGN, a pulsar or a
white dwarf, for the effect to be non-negligible. Furthermore we
need the ``chemistry'' of such an object to be sufficiently
simple, so that line blending does not become problematic.

\section{A last exercise for the student: MOND}
It is sadly the case that we can't finish this review with
resounding conclusions. Rather we will peter out in the realm of
uncertainties -- problems which remain unsolved and that may have
something to say about phenomenological quantum gravity. For this
purpose we have selected the problem of dark matter in the
Universe.

Galactic rotation curves have long puzzled cosmologists.
Newtonian theory predicts that they should fall out like
$v_r\propto 1/r^{1/2}$. This follows from the simple
calculation:
\be
F=ma \rightarrow  {Mm\over r^2}=m{v^2\over r}\rightarrow v^2={M\over r}
\ee
Instead we observe them flattening out: $v\rightarrow v_\infty$.
A simple solution is that besides the visible matter there is
a halo of dark matter which dominates gravity on the outskirts
of the galaxy. This halo has the property $M_{DM} = A r$,
i.e. it must have a density profile $\rho_{DM} = B/ r^2$.
Thus
\be
{M_{DM}m\over r^2}=m{v^2\over r}\rightarrow  v^2=A
\ee
Historically this the first hint of dark matter
in the Universe. It is important to stress that there are now
many other reasons to invoke dark matter\footnote{It's not obvious that
the required matter is always the same.}.

This is all very well; however three difficulties are quickly encountered:
\begin{itemize}
\item{1. }
The halos don't appear to be stable when left to evolve according
to their own gravity. Rather they collapse into a central cusp.
This is the drama of every N-body simulation performed so far.
Lack of resolution and physical content is usually blamed.
\item{2. } The onset of the terminal velocity seems to be triggered
not by a length or mass scale but by an acceleration. This
has been measured to be $a_0\approx 10^{-10}ms^{-2}$.
\item{3. } An empirical law has been established called
the Tully-Fisher relation
establishing that $v_\infty^4$ is proportional to the luminosity
(which presumably is proportional to the visible mass). This
is the equivalent of Kepler's third law.
\end{itemize}
It is hard to see how dark matter, even if creating a stable
halo, could explain the Tully-Fisher relation. There would have
to be a finely tuned  correlation between constant $B$ (appearing in the
density profile for the halo)  and the mass in visible matter.
Likewise the emergence of $a_0$ in the dark matter scenario is
hard to understand.
Still, it is possible that future N-body simulations may solve
these problems.

Disconcertingly there is a very simple alternative solution, called
MOND (MOdified Newtonian Dynamics). Perhaps galactic rotation
curves are simply telling us that gravity has departed from
Newton's equations (and that there is no
dark matter). Changing Newton's gravitational law, however, won't do because
this would trigger novel behaviour at a given length scale
rather than at an acceleration scale. Instead MOND
posits that {\it the response law}, $F=ma$, must be modified.
The usual law is only valid at high accelerations; for accelerations
smaller than $a_0$ we have instead that
\be
F=m{a^2\over a_0}\,  .
\ee
Straightforward application of the MOND prescription leads to
\be
 {Mm\over r^2}=m{v^4\over a_0 r^2}\rightarrow v^4=M a_0
\ee
Thus the Tully-Fisher relation is trivially explained  as well as the fact
that novel behaviour is triggered by an acceleration.

MOND is an excellent phenomenological description of galactic
rotation curves. However it makes no sense whatsoever. Applied
crudely it violates energy and momentum conservation in ways that
would readily conflict with observations. It also has
no relativistic generalization; we need such a relativistic
theory in order, for consistency,  to do away with
dark matter in all regimes. We quote the exemples of
gravitational lenses, the cosmological expansion, and structure
formation. Again, we're left in the
realm of wishful thinking: for dark matter  with regards to
computer power, here with
respect to brain power and essential theoretical developments.

Why might this discussion be relevant to quantum gravity?
Most obviously because MOND leads to a scary possibility:
{\it in trying to quantize gravity we may have chosen the wrong classical
theory}. No wonder we're stuck. Our failures could simply
signal that we don't yet have the correct classical theory of gravity.
This is a speculation, but one with dramatic, far-reaching consequences.
As the title of this section shows, whatever one makes of it,
this is very much exercise for the student~\footnote{Warning:
this is an exercise
for a very keen
graduate student hoping to start a high risk, high return project.}.
Specifically, here's the problem:
\begin{itemize}
\item{1.} Find a relativistic version of MOND consistent with
energy and momentum conservation and capable of explaining
gravitational lenses, and all the successes of the Big Bang
without dark matter. \item{2.} Quantize this theory.
\end {itemize}
The following points may be interesting hints (but then again,
they may also be red herrings):
\begin{itemize}
\item The Pioneer puzzle, related to the anomalous acceleration
suffered by these satellites in their courses outside the Solar system,
is associated with acceleration $a_P=8.74\pm 1.33\times 10^{-10}m s^{-2}$.
\item The observed cosmic acceleration is of the same order
as $a_0$.
\item In contradiction with Mach's principle there appears
to be absolute frames in the Universe for acceleration
(but not speed). Why?
\end {itemize}
We'll leave a single reference on this topic:~\cite{bekmond}.




\bibliographystyle{aipproc}   


\bibliography{sample}

\IfFileExists{\jobname.bbl}{}
 {\typeout{}
  \typeout{******************************************}
  \typeout{** Please run "bibtex \jobname" to optain}
  \typeout{** the bibliography and then re-run LaTeX}
  \typeout{** twice to fix the references!}
  \typeout{******************************************}
  \typeout{}
 }

\end{document}